\documentclass[preprint,journal]{vgtc}       

\ifpdf
  \pdfoutput=1\relax                   
  \usepackage{graphicx}                
  \DeclareGraphicsExtensions{.pdf,.png,.jpg,.jpeg} 
\else
  \usepackage{graphicx}                
  \DeclareGraphicsExtensions{.eps}     
\fi%

\graphicspath{{figures/}{pictures/}{images/}{./}} 

\usepackage{microtype}                 
\PassOptionsToPackage{warn}{textcomp}  
\usepackage{textcomp}                  
\usepackage{mathptmx}                  
\usepackage{times}                     
\usepackage{cite}                      
\usepackage{tabu}                      
\usepackage{booktabs}                  
\usepackage{paralist}

\usepackage{xcolor}
\usepackage{balance}
\usepackage{enumitem}
\usepackage{hyperref}
\usepackage{setspace}

\newcommand{\systemname}{ACSeeker} 
\newcommand{\groupviewname}{Factor View}
\newcommand{\comparisonviewname}{Cluster View}
\newcommand{\personviewname}{Person View}
\newcommand{\horizonchartname}{Horizon Chart Group}
\newcommand{\matrixchartname}{Impact Timeline} 
\newcommand{\navigationtimelinename}{Navigator}
\newcommand{\matrixtimelinename}{MatrixLine}
\newcommand{\careerlinename}{CareerLine}
\newcommand{\wyf}[1]{\textcolor{black}{#1}}
\newcommand{\wyfroundtwo}[1]{\textcolor{black}{#1}}
\newcommand{\etal}{et al.}
\newcommand{\ie}{i.e.}
\newcommand{\eg}{e.g.} 
\newcommand{\ea}{$E_{A}$} 
\newcommand{\visa}{$P_{A}$} 
\newcommand{\visb}{$P_{B}$} 
\newcommand{\visc}{$P_{C}$} 
\newcommand{\visd}{$P_{D}$} 

\onlineid{1226}

\vgtccategory{Research}
\vgtcpapertype{application/design study}

\title{Seek for Success: A Visualization Approach for Understanding the Dynamics of Academic Careers} 

\author{Yifang Wang, Tai-Quan Peng, Huihua Lu, Haoren Wang, Xiao Xie, Huamin Qu, and Yingcai Wu} 
\authorfooter{
\item
    Yifang Wang was with the State Key Lab of CAD\&CG, Zhejiang University and the Hong Kong University of Science and Technology. A part of this work was done when she was a visiting student supervised by Yingcai Wu at Zhejiang University. E-mail: yifang.wang@connect.ust.hk. 
\item
    Tai-Quan Peng is with Michigan State University. E-mail: pengtaiq@msu.edu and winsonpeng@gmail.com. 
\item
    Huihua Lu, Haoren Wang, and Yingcai Wu are with the State Key Lab of CAD\&CG and Xiao Xie is with the Department of Sport Science, Zhejiang University. Yingcai Wu is the corresponding author. E-mail: \{huihualu, haorenwang, xxie, ycwu\}@zju.edu.cn. 
\item
    Huamin Qu is with the Hong Kong University of Science and Technology. E-mail: huamin@cse.ust.hk. 
}

\shortauthortitle{Biv \MakeLowercase{\textit{et al.}}: Global Illumination for Fun and Profit}

\abstract{
How to achieve academic career success has been a long-standing research question in social science research. 
With the growing availability of large-scale well-documented academic profiles and career trajectories, scholarly interest in career success has been reinvigorated, which has emerged to be an active research domain called the Science of Science (i.e., SciSci). 
In this study, we adopt an innovative dynamic perspective to examine how individual and social factors will influence career success over time. 
We propose \textit{\systemname}, an interactive visual analytics approach to explore the potential factors of success and how the influence of multiple factors changes at different stages of academic careers. 
We first applied a Multi-factor Impact Analysis framework to estimate the effect of different factors on academic career success over time. 
We then developed a visual analytics system to understand the dynamic effects interactively. 
A novel timeline is designed to reveal and compare the factor impacts based on the whole population. 
A customized career line showing the individual career development is provided to allow a detailed inspection. 
To validate the effectiveness and usability of \textit{\systemname}, we report two case studies and interviews with a social scientist and general researchers. 
} 


\CCScatlist{ 
 \CCScat{K.6.1}{Management of Computing and Information Systems}%
{Project and People Management}{Life Cycle};
 \CCScat{K.7.m}{The Computing Profession}{Miscellaneous}{Ethics}
}

\begin{document}

\begin{spacing}{0.97}

\firstsection{Introduction}
\label{sec:01_Introduction}
\maketitle


How to achieve individual career success is a long-standing research question that has been studied in various social science disciplines, such as sociology, organizational behaviors, and information science. 
With the increased availability of academic profiles such as researchers' careers and scientific outputs, 
academic careers have become one of the prominent topics in the study of careers that attracts the attention of both social scientists and general researchers. 
Social scientists want to unravel factors that will positively or negatively contribute to academic career success. 
Researchers in other disciplines are concerned with how to raise scientific productivity and achieve career success. 
This line of research has regained its prominence with the emergence of Science of Science~\cite{fortunato2018science} in the age of computational social science~\cite{lazer2009life}. 

Previous studies have focused on career success in terms of an individual's position or promotions within an institution. 
However, in the current boundaryless career world~\cite{arthur2005career}, it is not unusual for researchers to take a more flexible approach to pursue their career success across institutions and even social sectors. 
%
This change poses new conceptual and methodological demands for empirical research on academic career success. 
The first demand is a new perspective to capture the sequential patterns in researchers' career paths. 
The traditional point-to-point transition perspective fails to capture the long-term impact of historical events on the upcoming career performances. 
The classical time series modeling assumes only quantitative changes in a career path, making it challenging to capture qualitative changes. 
The second demand is a more panoramic framework to identify potential factors contributing to academic career success. 
Previous studies have well documented the impacts of individual factors (\eg, educations). 
Nevertheless, there is no doubt that researchers' careers are also contingent on their social connections (\eg, collaborations). 
The third demand lies in a more computationally efficient way to detect driving factors underlying academic career success and an informative way to analyze the intricate and dynamic relationships between factors and academic career success. 

\wyf{By harvesting multiple data sources in the visualization field as a context for illustration,} the current study adopts a dynamic perspective to understand how individual and social factors will influence researchers' career success by using an interactive visual analytics approach. 
However, developing such a system faces three challenges. 
\wyf{First, distilling the potential factors on academic career success and analyzing their dynamic impacts over time are difficult. 
In addition to individual factors, operationalizing social factors (\eg, collaborations) aggravates the complexity of the problem given their network nature, let alone organizing these factors into proper temporal formats. 
Moreover, capturing the effects of these factors requires a dynamic multivariate analytical framework given the long period of research fields.} 
\wyf{Second, visually presenting the effects of multiple factors over time is challenging. 
Such effects are always with complex data structures in social science such as multi-dimensional, temporal, or with pairwise comparisons. 
Visual data representations require supporting both cross-sectional and longitudinal studies of factors.} 
\wyf{Third, supporting the exploration of both rich academic profiles and the multi-factor effects on academic career success is non-trivial. 
Experts desiderate to combine the impacts and patterns with the academic profiles to comprehensively understand the results~\cite{shneiderman2003eyes}. 
Statistical summaries and coordination among diverse data aspects could be challenging given multiple sources and dimensions.} 

\wyf{To address the first challenge, we identify a set of individual and social factors based on social theories and our domain expert}. 
We then propose a novel framework to analyze multi-factor effects on academic career success over time. 
\wyf{For the second challenge}, we design novel visualization to reveal the above time-varying effects. 
Specifically, the \textit{\matrixchartname} is for comparing the effects of different categories within a factor. 
A \textit{\careerlinename} shows one's academic career development with multiple factors. 
\wyf{We solve the third challenge} by proposing \textit{\systemname}, an interactive visualization system that assists social scientists in exploring academic career success with multiple factors from different levels of detail. 
Our contributions are listed as follows. 
\begin{compactitem}
    \item We characterize the problem domain of visual analytics of time-varying effects of multiple factors on academic career success. 
    \item We propose a novel framework that analyzes the effects of multiple factors on researchers' career success longitudinally. 
    \item We develop \textit{\systemname}, an interactive visualization system for \wyf{social scientists} and other scholars to explore academic careers with multiple factors through novel designs. 
    \item We demonstrate the effectiveness and usability of \textit{\systemname} with a dataset \wyf{that involves more than 1,100 visualization researchers}. 
\end{compactitem}

\section{Related Work}
\label{sec:02_RelatedWork}

In this section, we discuss related work in both career data analysis and sequence data analysis. 

\subsection{Career Data Analysis and Visualization}
\label{sec:02_RelatedWork_CareerAnalysis} 
Careers have been widely studied in social science and data science. 
%
Most studies focus on the common patterns or similarity comparisons of career paths in terms of job-related attributes, such as title ranks and organizations. 
Traditional social science studies focus on a macro-level analysis, which uses cross-sectional analysis to learn inflow and outflows across occupations~\cite{rosenbaum1979organizational, hout1983analysing}. 
Such point-to-point transition analysis fails to capture the careers from a long-term perspective to learn the impact of historical events on upcoming career performance. 
Works in data mining study from a micro-level to compare individual careers~\cite{xu2014modeling} and predict future jobs~\cite{xu2015learning, meng2019hierarchical}. 
It lacks a macro summary of careers, which is essential to understand the aggregated behavioral patterns in social science. 
%
Visualization studies use career data as a scenario in the event sequence analysis, including similarity analysis~\cite{janicke2015interactive, du2016eventaction, filipov2019cv3, wang2021interactive} and visual sequence summarization~\cite{guo2017eventthread, guo2018visual}. 
CV3~\cite{filipov2019cv3} is a system for comparing multiple resumes that assist recruiters in finding suitable candidates. 
Du~\etal~\cite{du2016eventaction} and Jänicke~\etal~\cite{janicke2015interactive} developed two systems respectively to recommend similar career paths of students or musicians for a target individual. 
Instead of clustering the whole sequences, EventThread~\cite{guo2017eventthread} and ET2~\cite{guo2018visual} use a fine-grained clustering to summarize career paths into latent stages. 

The increased availability of academic profile data (e.g., Aminer~\cite{tang2008arnetminer} and Vispubdata~\cite{isenberg2016vispubdata}) has called for a renovated perspective to study academic careers. 
Existing works analyze these data from multiple perspectives such as publication, citation, and collaboration networks. 
Egoslider~\cite{wu2015egoslider} and Egolines~\cite{zhao2016egocentric} distill the evolutionary collaboration networks to learn the academic interactions. 
Fung~\etal designed a tree metaphor to visualize one's publications~\cite{fung2016design}. 
VIS Author Profiles utilizes a template-based natural language generation to present a researcher's profiles. 
Other works use novel designs (e.g., linked matrix and flower metaphor) to show the scientific influence of different research entities (\eg, researchers and publications)~\cite{shin2019influence, wang2018visualizing}. 

However, most studies focus on career data, lacking the analysis of potential individual and social factors that can contribute to career success. 
A few visualization systems contain social factors (\eg, social networks) in career analysis. 
In the Interactive Chart of Biography~\cite{khulusi2019interactive}, institutional and denominational ties are distilled to show the relationships of musicologists. 
ResumeVis~\cite{zhang2018resumevis} contains an ego-network-based graph that summarizes the co-working relations of the target individual. 
Nevertheless, these social networks have not been correlated with career success. 
In this work, we use multiple sources of academic profiles and propose a new visual analytics framework to understand how multi-factors affect academic career success from a dynamic perspective. 

\subsection{Sequence Mining and Visualization}
\label{sec:02_RelatedWork_SequenceAnalysis}
Event sequences are continually studied in the past decades in both social science and computer science with many shared research concerns. 
Since our focuses are analyzing the dynamic impacts of multiple factors on career success, we would discuss the most relevant works in both sequence mining and visualization. 

Sequence mining in both disciplines can be summarized into three categories: pattern discovery~\cite{wu2020towards}, sequence inference~\cite{xie2020visual}, and sequence modeling~\cite{wang2019tac}, based on the survey of Guo~\etal~\cite{guo2020survey} and Piccarreta~\cite{piccarreta2019holistic}. 
Sequence clustering is a universal technique to extract common sequential patterns using unsupervised learning. 
Xu~\etal~\cite{xu2005survey} summarized different clustering strategies into three categories: proximity-based~\cite{xu2005survey, ritschard2018sequence, piccarreta2019holistic}, feature-based~\cite{guralnik2001scalable}, and model-based~\cite{zeng2019emoco, ming2019protosteer}. 
Specifically, proximity-based methods use the distance matrix to measure the similarity of sequences. 
Analysts first define and compute the pairwise dissimilarities between sequences and use clustering approaches to obtain sequential patterns. 
A critical step in proximity-based methods is to construct the distance matrix. 
Many sequence dissimilarity measures are proposed~\cite{studer2016matters, xu2005survey} such as Euclidean distance and Levenshtein distance. 
Besides clustering on the entire sequences, several studies propose to use stage analysis to represent the sequence progression with more details~\cite{guo2017eventthread, guo2018visual}. 
In addition to clustering, sequence inference (also noted as event history analysis in sociology) is another enduring topic. 
It estimates the influence of historical events or trajectories on upcoming events. 
Graphical models~\cite{bhattacharjya2020event} and regression analysis~\cite{stopar2018streamstory} are widely used. 
We enhance the approach by Rossignon~\etal~\cite{rossignon2018sequence} which combines sequence clustering with sequence inference to estimate the time-varying impacts of multiple factors on the upcoming career performance. 

A variety of visualization techniques are also developed to analyze sequential data. 
Guo~\etal~\cite{guo2020survey} have proposed a comprehensive survey of event sequence visual analytics approaches. 
Here we mainly summarize the most relevant visualization techniques in our work. 
A large number of designs adopt an intuitive horizontal timeline encoding. 
Flow-based visualization is a typical representation with great scalability to summarize large-scale sequence data, including tree-based~\cite{monroe2014interactive, liu2017coreflow} and Sankey-based~\cite{gotz2014decisionflow, wu2021tacticflow} structures. 
Two strategies are commonly used to reveal sequential patterns with fine granularity. 
One uses stage analysis to extract latent stages within the whole sequence to show progressions~\cite{guo2017eventthread, guo2018visual}. 
The other directly visualize original individual sequences as horizontal lines~\cite{xie2014vaet, chen2016gameflow, xie2020passvizor, deng2021visual}. 
Recent work by Bartolomeo~\etal~\cite{di2020sequence} further uses layered directed acyclic network together with layout optimization to align specific events among sequences. 
Similar visual representations are adopted in storyline visualization~\cite{tang2018istoryline, tang2020plotthread}, where each line represents an actor, and the vertical position encodes groups of actors based on different events. 
Another set of studies use matrix- or list-based techniques to visualize event sequences that are scalable~\cite{zhao2015matrixwave, du2016eventaction, wu2017ittvis, wang2021tac, wu2020visual}. 
However, most of the state-of-the-art studies cannot be directly used in the study of academic career success. 
First, they focus on pattern extraction while the effects of multiple factors are of great importance in our scenario. 
Second, they consider the relative time which aligns sequences to the same starting point, while absolute time is also important to learn different generations of researchers. 
We have proposed novel visual designs to fulfill the above requirements. 

\section{Background and System Overview}
\label{sec:03_Background}
In this section, we introduce the background and concepts used in our study, summarize the analytical tasks, and provide a system overview to demonstrate the whole pipeline. 


\subsection{Background and Concepts}
\label{sec:03_Background_Concepts} 
The study of academic career success has been an enduring topic in multiple fields such as SciSci~\cite{fortunato2018science} and \wyf{sequence analysis in social science}~\cite{ritschard2018sequence}. 
Learning the time-varying impacts of multiple factors on academic career success can benefit the understanding of typical career patterns for social scientists and individual career development for general researchers. 
We have worked closely with a social scientist ({\ea}) to solve this research problem. 
He has been conducting multi-factor analysis with sequential data in different social domains. 
By working with {\ea}, we summarized the following concepts to characterize the problem of the multi-factor effect on academic career success. 
\begin{compactitem}
    \item \textit{Career Success}, or \textit{Career Performance}, refers to the outcomes (\eg, title ranks and incomes) of one's working experiences~\cite{seibert2001social, arthur2005career}. 
    In academic careers, \textit{citations} of a scholar's research outputs are commonly used to measure career success~\cite{fortunato2018science}. 
    \item \textit{Career Factors} refer to potential factors (\ie, variates) that can affect \wyf{one's} career success. 
    It includes both individual (e.g., personal characteristics) and social factors (e.g., social relations)~\cite{zacher2019academic}. 
    In academic careers, due to the data accessibility, we collected four factors that may affect career success based on previous empirical studies and our expert's suggestions. 
    Job title ranks, sectors (\eg, academia, industry, and government agencies), and research domains are individual factors. 
    \wyf{As collaboration is an essential type of  occupation network in academia that can have significant effects on careers~\cite{fortunato2018science, arthur2005career}, we regard collaborators' individual factors as social factors~\cite{zacher2019academic}}. 
    \item \textit{Factor Categories} are different groups identified \wyf{based on the values within a factor according to user-specified definitions. 
    For example, users may define people within the same value range of a factor as one category.} 
\end{compactitem}

\subsection{Data Description}
\label{sec:03_Background_DataDescription} 
\wyf{The analysis of dynamic multi-factor effects on academic career success is based on data from multiple sources, some of which require complicated and even manual data preparation.} 
Considering the data availability and our familiarity, we focus on researchers from the visualization (\ie, VIS) community as a context to illustrate academic career analysis. 
We consider those who have published more than two TVCG papers in which the largest time gap is more than five years as potential VIS researchers \wyf{after discussing with our expert}. 
We then manually check and filter out those from other fields \wyf{(about 90 researchers)} and finally obtain over 1,100 VIS researchers. 
\wyf{We use researchers' names as inputs to search different data sources below, which are transformed into multiple sequences by each researcher for in-depth analysis.} 
\begin{compactitem}
    \item \textbf{Career data} of researchers record the job-related attributes such as the institutions and titles. We collected it from LinkedIn~\cite{linkedin}, researchers' personal websites, and their institutional webpages. 
    \item \textbf{Bibliographic data} is \wyf{directly} gathered from Aminer~\cite{tang2008arnetminer}, \wyf{which includes over 21,600 papers in total based on these researchers}. 
    It includes all the publication metadata of a researcher (e.g., authors, year, venue, title, and abstract) by year. 
    \item \textbf{Citation data} (by year) \wyf{is crawled} from Google Scholar~\cite{googlescholar} as a measure of career success~\cite{fortunato2018science}. 
\end{compactitem}

\subsection{Task Analysis}
\label{sec:03_Background_TaskAnalysis} 
Our goal is to analyze the dynamic effects of multiple factors on academic career success. 
Guided by the nested model of visualization design~\cite{munzner2009nested}, we conducted literature reviews and frequent interviews with the expert to iteratively distill and refine the tasks. 
Finally, we form the analytical tasks into three levels~\cite{bertin1983semiology} as follows. 

The \textbf{inter-factor-level} task provides an overview of how different factors contribute to career success. 
\begin{enumerate}[leftmargin=*,label={\textbf{T{\arabic*}}}]
    \setlength{\itemsep}{2pt}
    \setlength{\parskip}{2pt}
    \setlength{\parsep}{2pt}
    \vspace{-2mm}
    \item \textbf{How does a factor influence career success over time?} 
    For longitudinal comparison, the expert wishes to know how the impact of a specific factor on career success develops over time. 
    This could be related to the development of specific academic fields. 
    \item \textbf{\wyf{How} do multiple factors differ in their impacts on career success?} 
    The expert \wyf{wants to know} the effects of different factors at a specific time as a cross-sectional comparison to determine the dominant factors influencing career success. 
    Specifically, it is interesting to compare the impacts of individual and social factors. 
\end{enumerate}

The \textbf{intra-factor-level} tasks drive into one specific factor to compare the change of effects of different categories on career success over time. 
\begin{enumerate}[leftmargin=*,label={\textbf{T{\arabic*}}}]
    \setlength{\itemsep}{2pt}
    \setlength{\parskip}{2pt}
    \setlength{\parsep}{2pt}
    \setcounter{enumi}{2}
    \vspace{-2mm}
    \item \textbf{How \wyf{does} a category within a factor \wyf{change} over time to affect career success?} 
    The impact of a category can change at different periods. 
    It reflects the change of the roles of this category. 
    \item \textbf{How \wyf{do} the categories within a factor differ from affecting career success?} 
    For each factor, different categories may have different sizes of impacts. 
    Figuring out those with high impacts can help explain and provide guidelines for academic career development. 
\end{enumerate}

The \textbf{individual-level} task gives several concrete examples to help understand the factor effects on career success. 
\begin{enumerate}[leftmargin=*,label={\textbf{T{\arabic*}}}]
    \setlength{\itemsep}{2pt}
    \setlength{\parskip}{2pt}
    \setlength{\parsep}{2pt}
    \setcounter{enumi}{4}
    \vspace{-2mm}
    \item \textbf{How \wyf{does a researcher's career path} change over time?} 
    The expert wishes to identify different individuals from the data and study the multi-factor effects on career success from a micro perspective.  
    \wyf{Revealing the career changes of a researcher over time} can help understand the different academic stages he goes over. 
    \item \textbf{\wyfroundtwo{How \wyf{do} the different factors of a researcher change over time?}} 
    \wyf{Showing the evolution of career factors over time is essential to interpret the \wyf{career development} of a researcher and further validate existing rules or generate new hypotheses.} 
\end{enumerate}

\subsection{System Overview}
\label{sec:03_Background_SystemOverview} 
\begin{figure} [!htb]
 \centering 
 \vspace{-0.2cm}
 \includegraphics[width=\linewidth]{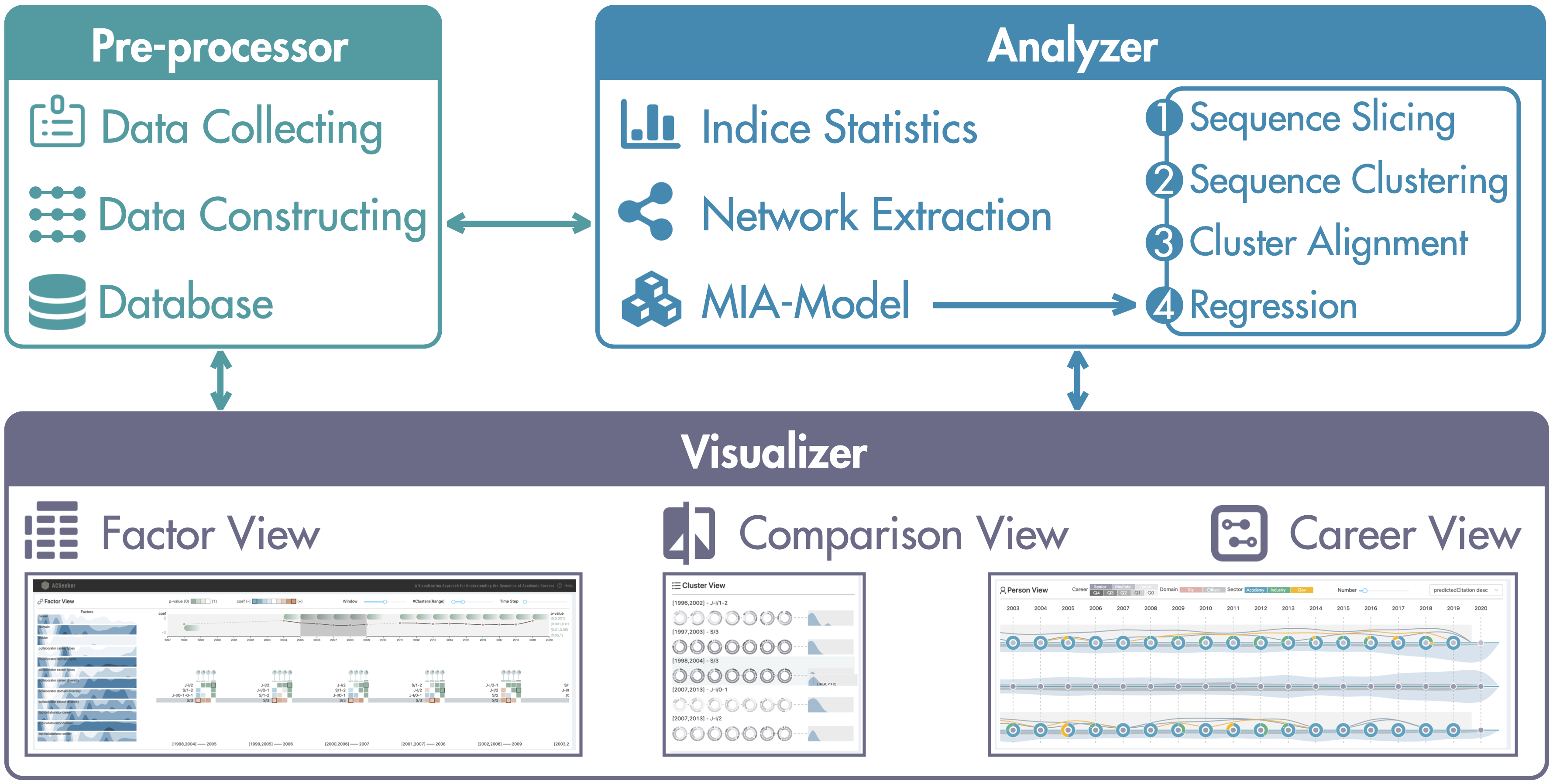}
 \vspace{-0.5cm}
 \caption{System overview. \textit{\systemname} has three components: a data pre-processor module, an analyzer module, and a visualizer module.} 
 \vspace{-0.3cm}
 \label{fig:system-overview}
\end{figure}

\textit{\systemname} is a web-based application with three modules: a data preprocessing module, a data analysis module, and a data visualization module (Fig.~\ref{fig:system-overview}). 
The data preprocessing module collects and cleans researchers' career profile, publication, and citation data. 
Then it organizes these data into multiple sequences and stores them in the database. 
The data analysis module utilizes a novel framework to analyze the effect of multiple factors on career success. 
They form into the backend of the system and are implemented using Python and MongoDB. 
The data visualization module constructs a frontend application using Vue.js~\cite{vue} and D3.js~\cite{bostock2011d3} with three views to support analysis. 

\section{Data Analysis}
\label{sec:04_Algorithm}
In this section, we first introduce a set of data preprocessing measures. 
\wyf{Then we introduce the existing Sequence History Analysis \wyfroundtwo{(SHA)}~\cite{rossignon2018sequence} applied in previous social science research. 
Finally, we describe our new analytical framework based on SHA to simultaneously extract career patterns and analyze the impacts of multiple types of historical sequences on career success over a long period.} 

\subsection{Data Preprocessing}
\label{sec:04_Algorithm_DataPreprocessing} 
\begin{figure*} [!htb]
 \centering 
 \vspace{-0.5cm}
 \includegraphics[width=\linewidth]{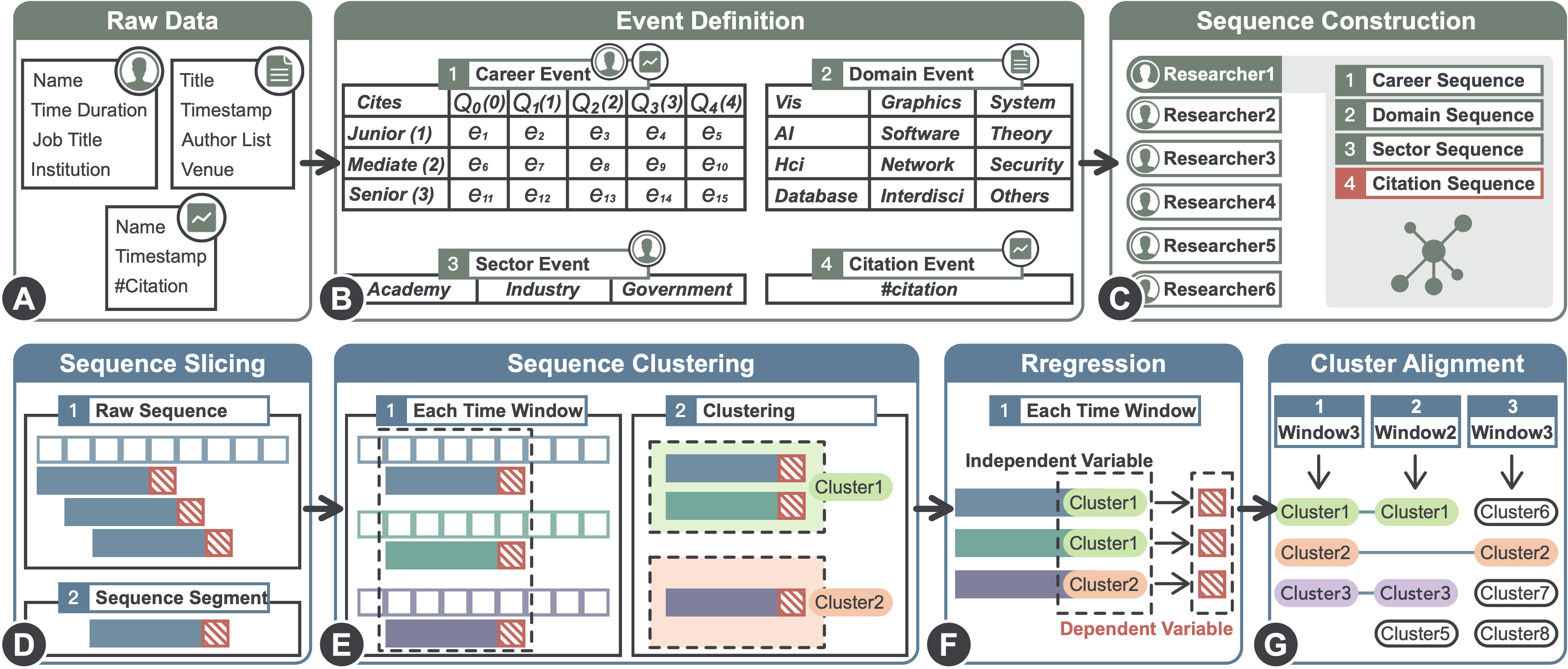}
 \vspace{-0.45cm} 
 \caption{The pipeline of data preprocessing and the multi-factor impact analysis (MIA) framework. 
 For data preprocessing (A-C), we first define the events based on raw data collected from three sources. 
 Then we construct four sequences for each researcher as the input of the framework. 
 The MIA framework (D-G) consists of four steps: 
 (D) Sequence Slicing,  
 (E) Sequence Clustering, 
 (F) Regression, 
 and (G) Cluster Alignment.} 
 \vspace{-0.5cm} 
 \label{fig:framework}
\end{figure*}

\wyf{After collecting multiple data sources (Section~\ref{sec:03_Background_DataDescription}), 
\wyfroundtwo{we preprocessed the data in a semi-automatic way}.} 
For the career data, we organized each job into an event with a timestamp (by year) and an institution. 
\wyf{We manually tagged the job titles and sectors in career data.} 
Job titles were tagged into three ranks (i.e., junior, intermediate, and senior) based on researchers' tenure in academic research (Fig.~\ref{fig:framework}-B1). 
We also tagged three sectors: academia, industry, and government agency, based on the institutions (Fig.~\ref{fig:framework}-B3). 
From the bibliographic data, we extracted the paper venues by year and classified them into twelve categories to represent different research domains based on~\cite{ccfstandard} (Fig.~\ref{fig:framework}-B2). 
We also extracted all the collaborators of a researcher by year to construct his ego-networks. 
For the citation data, we used Quartile~\cite{quartile} to divide each year's citations into four ranks of equal size (Fig.~\ref{fig:framework}-B1). 
In addition, we separated the top 3\% citation researchers from the top 25\% to inspect the pioneers in the visualization field. 

Finally, multiple sequences are constructed for each researcher (Fig.~\ref{fig:framework}-C). 
\textit{Career Sequences} are composed of events defined by both title and citation ranks ordered by year. 
\textit{Domain Sequences} consist of a list of paper venue categories with corresponding numbers of papers each year. 
\textit{Sector Sequences} are sequences of sectors where researchers are affiliated over time. 
\textit{Citation Sequences} are sequences of citation numbers by year. 
Meanwhile, we record all the collaborators of each researcher by year.
We use their career, domain, and sector sequences as social factors. 
It allows us to quantify how a researcher's collaborators can influence his career success. 
These sequences, along with the collaboration networks, are fed into our analytical framework.


\subsection{\wyf{Sequence History Analysis (SHA)}}
\label{sec:04_Algorithm_SHA} 
\wyf{A body of literature in social science has studied the effect of historical event trajectories on an upcoming target event~\cite{madero2016influence}. 
Most of them reduce the historical sequences into summary indicators (e.g., event duration)~\cite{blossfeld2007event}, which fail to keep the full information in the original sequences. 
Sequence History Analysis \wyfroundtwo{(SHA)}~\cite{rossignon2018sequence} is an innovative approach to preserve more complex sequential information in two steps. 
First, \textit{Sequence Analysis}~\cite{ritschard2018sequence} is applied to identify representative patterns over the historical sequences. 
A distance matrix is constructed to document the pairwise distances between raw sequences. 
Using this matrix, the sequences are clustered into groups (\ie, categories) based on clustering algorithms (e.g., k-means). 
It retains the most representative sequential patterns over raw sequences and substantially reduces the computational demand in the subsequent multivariate analysis. 
Second, \textit{Event History Analysis}~\cite{blossfeld2007event} is used to analyze how these historical sequential patterns will affect the upcoming event. 
Different regression models will be applied to obtain the estimation of the effects.} 

\wyf{However, the original SHA cannot be directly adopted in our scenario. 
First, it analyzes the impacts in a static manner by aligning sequences to the same starting point. 
However, the impacts of historical trajectories of factors could vary over time due to the development of the research field. 
Second, 
the SHA is supposed to analyze only one type of sequences, 
while in our scenario, multiple types of sequences are hypothesized to influence career success. 
}

\subsection{Multi-factor Impact Analysis (MIA)}
\label{sec:04_Algorithm_MIA} 
\wyf{We have worked with our domain expert to} enhance the SHA approach to \wyf{support dynamic analysis of the impacts of multiple factors} on academic careers over time. 
The whole framework consists of four components: sequence slicing, sequence clustering, multivariate linear regression, and cluster alignment (Fig.~\ref{fig:framework}-D, E, F, G). 

\textbf{Sequence Slicing.} 
\wyf{We first improve SHA by slicing multiple sequences into different time windows.} 
\wyf{Given a long period, our expert hoped to inspect the time-varying impacts of different factors. 
Moreover, he stated the importance of considering the time-sensitivity of the impacts of historical sequences: the farther the historical event, the less relevant it is to the upcoming career performance. 
We thus apply the sliding window method~\cite{chu1995time, yu2014time} to meet the two requirements.} 
Each researcher's career-related sequences (Fig.~\ref{fig:framework}-D1) are arranged along the absolute timeline and sliced into different windows of a fixed size (i.e., size $w$, Fig.~\ref{fig:framework}-D2). 
We aim to analyze the impacts of multiple factors in each time window separately. 
\wyf{The window size and the moving step can be adjusted based on domain knowledge in \textit{\systemname}.} 

\textbf{Sequence Clustering.} 
\wyf{In each time window, we follow the \textit{Sequence Analysis} method in the SHA approach to identify the most representative sequential patterns of each factor. 
It uses sequence clustering on each type of sequences (i.e., career, domain, and sector) respectively (Fig.~\ref{fig:framework}-E).} 
%
After comparing different clustering algorithms (e.g., k-means, k-medoids, agglomerative, DBSCAN, optics, and spectral), we finally chose k-means due to its optimal performance and efficiency. 
\wyf{We use Euclidean distance to obtain the dissimilarity matrix, which is widely applied in social science with remarkable computational efficiency~\cite{studer2016matters}.} 
\wyf{As our expert expected to adjust the number of clusters to obtain more meaningful results, the system allows users to customize the range of cluster numbers in k-means.} 

\textbf{Multivariate Linear Regression.} 
\wyf{In each time window ($[t, t+w]$) (Fig.~\ref{fig:framework}-F), we then conduct a multi-factor impact analysis on career success. 
We improve the SHA approach by applying ordinary-least-square (OLS) regression to incorporate multiple factors.} 
\begin{equation}
\vspace{-1mm}
    y_{i} = \beta_{0} + \beta_{1}x_{i1} + \beta_{2}x_{i2} + ... + \beta_{n}x_{in} + \varepsilon_{i},  (i=1,...,n)
\end{equation}
We have incorporated twelve independent variables (i.e., IVs, including individual and social factors) and a dependent variable (i.e., DV) based on our expert's suggestions as follows: 
\begin{compactitem}
    \vspace{-1mm}
    \item \textbf{$IV_{1}-IV_{6}$}. 
    Six categorical IVs are based on the clustering results of three types of historical sequences (\ie, \textit{Career}, \textit{Domain}, and \textit{Sector Sequences}) of researchers ($IV_{1}-IV_{3}$) and their top collaborators ($IV_{4}-IV_{6}$) who have the most co-authored papers in a time window. 
    \wyf{We choose the top collaborator as a representative that potentially affects one's career significantly (\eg, advisors or long-term collaborators). 
    To include categorical variables as IVs in OLS regression, dummy coding~\cite{hardy1993regression, dummyvariable} is employed, transforming a categorical IV with $n$ categories into $n-1$ \textit{dummy variables}~\cite{dummyvariablewiki}. 
    Specifically, an arbitrary category of a categorical IV is chosen to serve as the \textit{reference category} and all other categories are set to be the \textit{comparison} or \textit{target categories}. 
    The obtained coefficient of each dummy variable means how a comparison category performs on the career success compared with the \textit{reference category}. 
    The obtained p-value shows the statistical significance of this comparison. 
    As the choice of the \textit{reference category} is arbitrary, we improve SHA by conducting a post-hoc analysis, which enables pairwise comparisons on the career success among all categories of a categorical IV as a panoramic view. 
    Specifically, we traverse to set each of the $n$ categories as a \textit{reference category} in turn in the regression. 
    It results in a matrix-like table recording the coefficients and p-values of pairwise comparisons.} 
    \item \textbf{$IV_{7}-IV_{12}$}. 
    \wyf{Besides the top collaborator, our expert suggested having a global summary of one's overall collaborators in the model. 
    We thus include two types of numerical variables. 
    The first type ($IV_{7}-IV_{9}$) is the number of collaborators in each sequence cluster weighted by the collaboration strength.} 
    It is the product of the proportion of collaborators of this cluster and the normalized number of papers he collaborated over all the papers he published within the time window. 
    The second type ($IV_{10}-IV_{12}$) counts the number of sequence clusters for each sequence type. 
    \item \textbf{$DV$}. 
    We use the number of citations at the following year of the time window to measure career success~\cite{fortunato2018science}. 
    We take the logarithm of the citations as DV since the raw citations each year are with a positively skewed distribution. 
\end{compactitem} 

\begin{figure*} [t]
 \centering 
 \includegraphics[width=\linewidth]{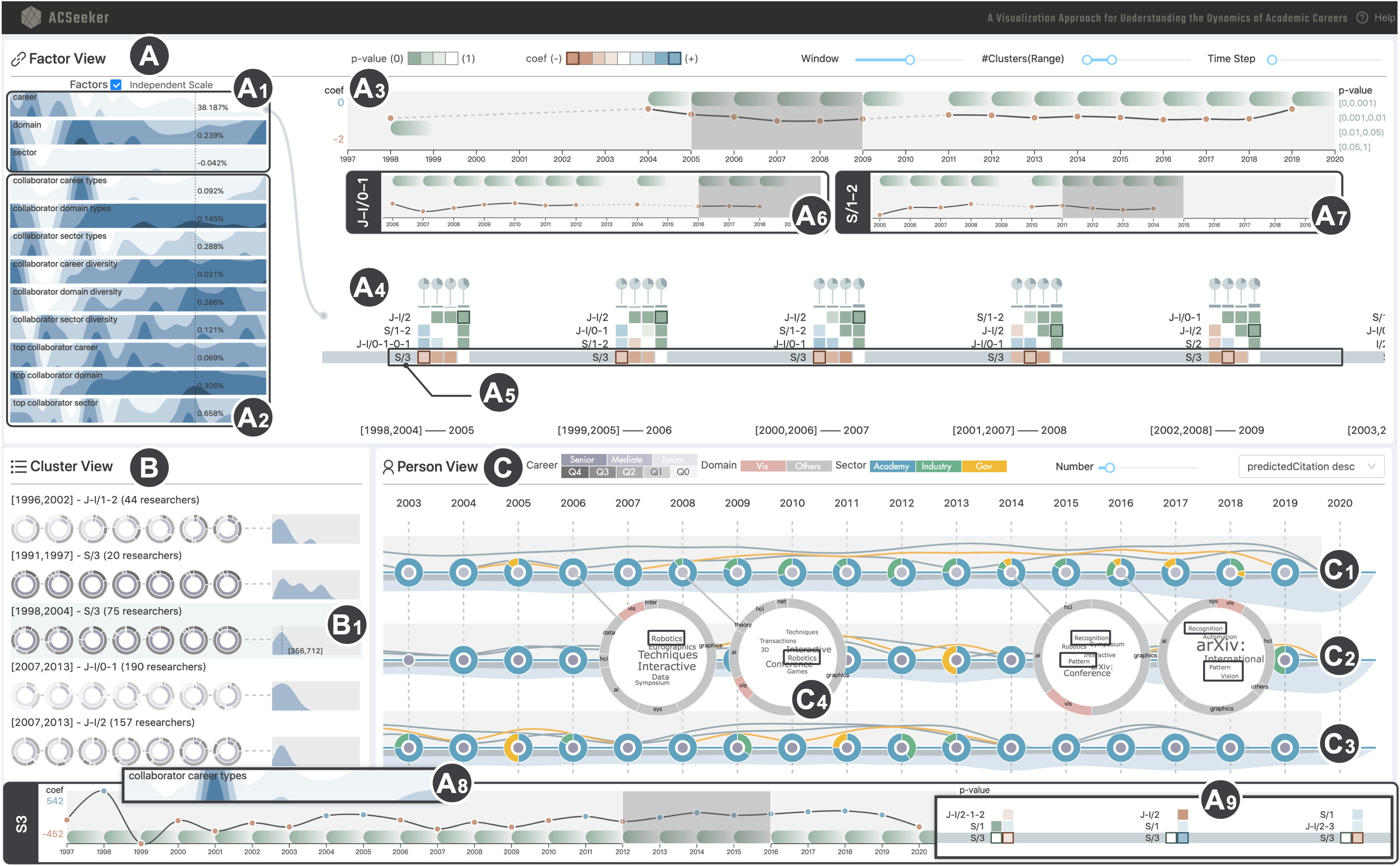} 
 \vspace{-0.5cm}
 \caption{The system interface of \textit{\systemname}. 
 (A) The \textit{\groupviewname} reveals the regression results. 
 It consists of three parts: \textit{\horizonchartname} for inter-factor analysis, \textit{\navigationtimelinename} and \textit{\matrixtimelinename} for intra-factor analysis. 
 (B) The \textit{\comparisonviewname} summarizes a list of sequence clusters chosen in the \textit{\groupviewname}. 
 (C) The \textit{\personviewname} provides detailed information of careers and multi-factors for each researcher using a novel \textit{\careerlinename} design.} 
 \vspace{-0.5cm}
 \label{fig:system-ui}
\end{figure*}

\wyf{In each window, we compute the explanatory power (\ie, impact) of each IV (\ie, factor), which is measured by the difference of $R^{2}$ for whether adding this IV into the regression.} 

\textbf{Cluster Alignment.} 
\wyf{In the last step, 
our expert wanted to align the same cluster in each sequence type across windows to learn its effect on career success temporally, which is not supported by SHA.} 
We adopt a naive approach to label clusters by the event with the largest proportion in the cluster by year \wyf{as it is the most representative one}. 
We regard those with the same label as the same cluster and conduct the alignment \wyf{across windows} using these labels (Fig.~\ref{fig:framework}-G). 

In addition, for each researcher, we compute three diversity scores (\ie, career, domain, and sector types) of his collaborators by year using entropy ($n$ is the number of total event types): 

\vspace{-4mm} 
\begin{equation}
    Diversity(k,t) = -\sum_{i=1}^{n}P(x_i,t)logP(x_i,t), k \in \{career, sector, domain\}
\end{equation}
 
The regression models of all windows (each includes \textit{p-values}, \textit{coefficients} and the difference of $R^{2}$ of each IV, and the predicted citations of each researcher) are fed into the \textit{\systemname} for further analysis. 

\section{Visual Design}
\label{sec:05_VisualDesign}
\wyf{\textit{\systemname} consists of three views to facilitate the analytical tasks in Section~\ref{sec:03_Background_TaskAnalysis}: the \textit{\groupviewname}, the \textit{\comparisonviewname}, and the \textit{\personviewname}. 
The workflow is as follows. 
\wyfroundtwo{Users will begin with the \textit{\horizonchartname} in the \textit{\groupviewname}} and choose a factor of interest for detailed analysis in the \textit{\matrixchartname}. 
Through \textit{\matrixtimelinename} and \textit{\navigationtimelinename}, based on domain knowledge, they may focus on the analysis of a specific category (\ie, sequence cluster) within a factor to learn its time-varying effect on career success. 
They may also adjust the number of clusters and the window size to obtain better results. 
During the exploration of \textit{\matrixtimelinename}, users can add interested clusters to the \textit{\comparisonviewname} for sequential comparisons. 
Finally, they can choose a cluster of interest in the \textit{\comparisonviewname} and use the \textit{\personviewname} to learn individuals' careers and factor impacts within the cluster through the \textit{\careerlinename} design. 
The rest of this section will follow the order of the workflow to introduce each visual component.
}

\subsection{\groupviewname}
The \textit{\groupviewname} (Fig.~\ref{fig:system-ui}-A) is the primary visual component to show the regression results. 
It consists of two parts: (1) the \textit{\horizonchartname} supports the inter-factor comparison on career success (\textbf{T1}, \textbf{T2}); (2) the \textit{\matrixchartname} allows the intra-factor inspection within a factor (\textbf{T3}, \textbf{T4}). 
Users can have an overview of the time-varying effects of multiple factors and choose groups of interest for further study. 

\subsubsection{Inter-Factor-Level Analysis} 
The \textit{\horizonchartname} (Fig.~\ref{fig:system-ui}-A1, A2) summarizes the trends of factor impacts and supports the comparison among factors (\textbf{T1}, \textbf{T2}). 

\textit{Description}: 
Each horizon chart represents a factor which is extended from a line chart. 
The x-axis encodes the time and the y-axis represents the explanatory power of a factor (\ie, $R^{2}$). 
The line chart is divided into layered bands with uniform ranges. 
The y value is encoded by a gradient color scheme in blue. 
The darker the color, the higher the value. 
Then the bands are shifted to the center and distributed within a fixed height. 
\wyf{Two y scales are provided: a unified scale for impact comparison across factors and an independent scale for temporal inspection within a factor. 
Users can choose a factor for further analysis.} 

\textit{Justification}: 
Initially, we used line charts with two modes.   
However, a multi-line graph that includes all the factors in one coordinate suffered great visual clutters. 
Small multiples where each line chart represented a factor were also not appropriate, since the height of each line chart was too narrow to show the temporal trends, let alone the comparison among factors. 
Thus, we chose the horizon chart to show the impact of multiple factors in a compact way. 

\subsubsection{Intra-Factor-Level Analysis} 
\begin{figure} [t]
 \centering 
 \vspace{-0.5cm}
 \includegraphics[width=\linewidth]{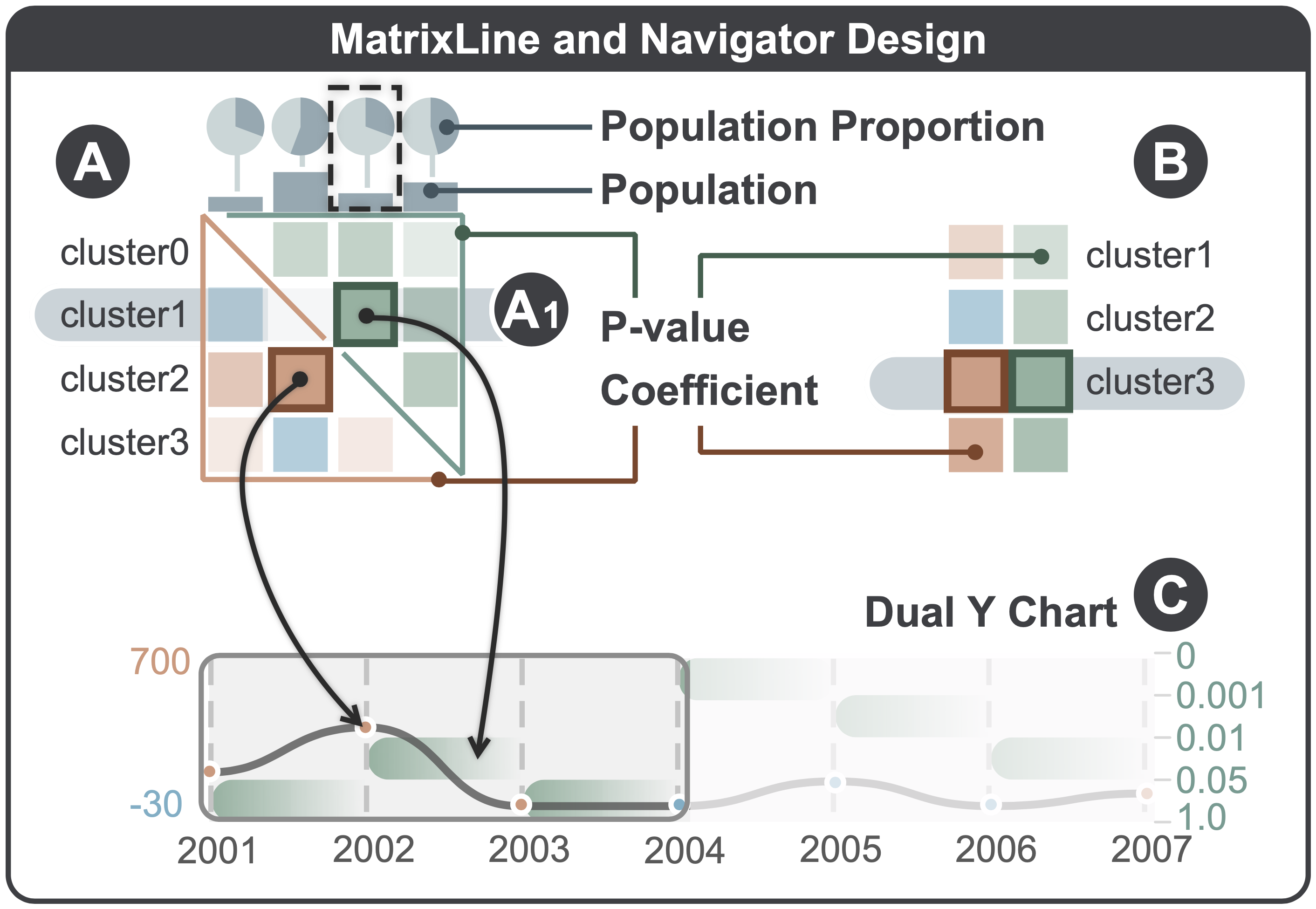}
 \vspace{-0.5cm} 
 \caption{The visual design of \textit{\matrixtimelinename} and \textit{\navigationtimelinename}.
 (A) The $n*n$ matrix design for categorical variables with $n$ clusters. 
 (B) The $n*2$ matrix design for numerical variables. 
 (C) The \wyf{dual-y} chart with coefficients and p-values obtained in the regression.}  
 \label{fig:visual-design-1-matrix}
 \vspace{-0.55cm}
\end{figure}

\wyf{After specifying the factor in the \textit{\horizonchartname}}, users can use the \textit{\matrixchartname} (Fig.~\ref{fig:visual-design-1-matrix}) to study the detailed impacts of the factor obtained from the regression model. 
Users can observe the temporal trends of the impact and compare the effects of different categories within the factor through intuitive interactions (\textbf{T3}, \textbf{T4}). 

\textit{Description}: 
The \textit{\matrixchartname} consists of two parts: a \textit{\matrixtimelinename} showing the detailed impact of a \wyf{time window} (Fig.~\ref{fig:visual-design-1-matrix}-A, B) and a \textit{\navigationtimelinename} revealing the \wyf{impact evolvement over time} (Fig.~\ref{fig:visual-design-1-matrix}-C). 

\textbf{\matrixtimelinename.} 
\wyf{Each matrix shows the MIA model results of a time window (\textbf{T4}). 
The design targets independent variables related to sequence clustering results (\ie, $IV_{1}-IV_{9}$ in Section~\ref{sec:04_Algorithm_MIA}).} 
\wyf{To show the impacts of different categories (\ie, clusters) in individual's ($IV_{1}-IV_{3}$) and his top collaborator's factors ($IV_{4}-IV_{6}$), our MIA model transforms these categories into \textit{dummy variables} and uses post-hoc analysis to produce pairwise category comparisons. 
The output is an $n*n$ matrix for coefficients and p-values in the regression as mentioned in Section~\ref{sec:04_Algorithm_MIA}.} 
Thus, in Fig.~\ref{fig:visual-design-1-matrix}-A, we divide the matrix into an upper triangle and a lower triangle to show the pairwise p-values and coefficients, respectively. 
In the upper triangle, each cell represents the pairwise p-value. 
The p-value is partitioned to four ranges (i.e., [0, 0.001), [0.001, 0.01), [0.01, 0.05), [0.05, 1]) based on \wyf{the traditional social science approaches~\cite{boos2011p, walsh2014statistical} and our expert's suggestions to show the statistical significance}. 
\wyf{We use white for range [0.05, 1] (\ie, not statistically significant) and green in different saturations for the other three ranges to distinguish the statistical significance at different levels.} 
The smaller the p-values, the darker the green. 
In the lower triangle, each cell shows the coefficient of pairwise categories. 
Our expert emphasized the importance of studying both the absolute values and the positive and negative of the coefficients. 
Thus, we use blue and red to encode the positive and negative values, respectively. 
The saturation encodes the absolute values. 
The larger the absolute value, the darker the color. 
\wyf{The bar charts and pie charts above the matrix summarize the population and proportions of individuals in each category (\ie, sequence cluster) respectively. 
The dark grey area in the pie chart represents the proportion of this cluster.} 

To inspect the time-varying impact of a category, users can first specify a \textit{reference category} (Fig.~\ref{fig:visual-design-1-matrix}-A1) to align all the matrices (Section~\ref{sec:04_Algorithm_MIA}) across time. 
It will skip matrices without the \textit{reference category}. 
Then they can choose a \textit{target category} by clicking its pie chart. 
Two cells showing the p-value and coefficient will be highlighted across time. 
\wyf{The \textit{\navigationtimelinename} will also be updated to summarize the temporal trends. 
The target category will be added to the \textit{\comparisonviewname} for further analysis.} 
\wyf{Users can customize the number of clusters, the length of the time window and the window moving step to adjust the MIA model.} 

\wyf{To show the impact of collaboration strength of each collaborator's category of a social factor (i.e., $IV_{7}-IV_{9}$)}, 
we transform the $n*n$ matrix into an $n*2$ one (Fig.~\ref{fig:visual-design-1-matrix}-B). 
Each row represents a category of this factor and two columns represent the p-value and coefficient respectively. 
Users can align a category to study its time-varying impact. 

\textbf{\navigationtimelinename.} 
\wyf{A \textit{\navigationtimelinename} provides a temporal overview of two values (\ie, p-value and coefficient) of a selected category mentioned above or the numerical independent variables (\ie, $IV_{10}-IV_{12}$) (\textbf{T3})}. 
It is a timeline with a dual y-axis that represents p-value and coefficient, respectively. 
\wyf{A stepped line graph} in green shows the p-values and the line chart in black reveals the coefficients. 
The color of the circle on the line chart encodes the positive and negative values with the same color scheme in the matrix. 
Users can quickly find the period of interest and drag to focus the time window in \textit{\matrixtimelinename}. 
For the numerical factors, users can directly use \textit{\navigationtimelinename} to study the time-varying impacts. 

\textit{Justification}: 
We designed two line charts for two values as the navigator while it was space-wasting. 
We then assembled them into \wyf{a dual-y chart} with an area chart encoding the p-value. 
\wyf{However, after trying the system, our expert suggested that the ranges of the p-values were more practical to learn the statistical significance than the raw values.} 
Thus, we adopted a stepped line graph to fulfill the requirement. 

\begin{figure} [t]
 \centering 
 \vspace{-0.5cm}
 \includegraphics[width=\linewidth]{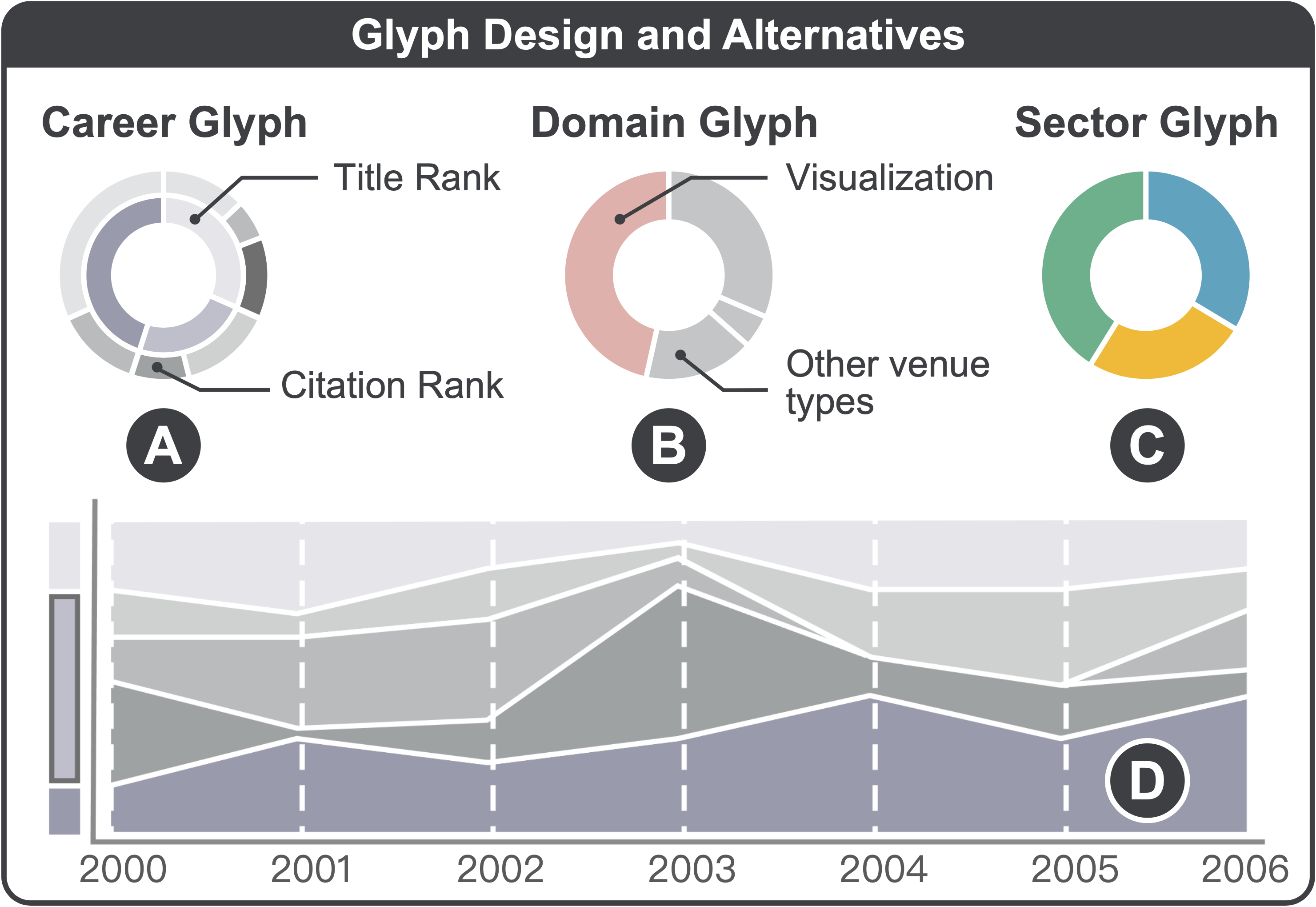} 
 \vspace{-0.5cm}
 \caption{Glyph designs and an alternative to summarize group information of careers and factors. 
 The career glyph (A) is a sunburst graph showing the title and citation ranks. 
 The domain (B) and sector (C) glyphs are donut charts showing the distribution of different categories of a group.} 
 \vspace{-0.55cm}
 \label{fig:visual-design-2-glyph}
\end{figure}

\subsection{\comparisonviewname}
\wyf{The \textit{\comparisonviewname} (Fig.~\ref{fig:system-ui}-B) presents a list of categories (\ie, sequence clusters) chosen from the \textit{\matrixchartname}.} 
Three glyphs (Fig.~\ref{fig:visual-design-2-glyph}-A, B, C) are designed to summarize the cluster of different sequences (\textbf{T4}). 

\textit{Description}: 
Each row shows a sequence cluster of a time window \wyf{with a title listing the window, the cluster label, and the number of researchers within the cluster.} 
Each glyph of the row summarizes the distribution of researchers within the cluster at one year. 
We have designed three glyphs to show the cluster summary of three types of sequences (Fig.~\ref{fig:visual-design-2-glyph}-A, B, C). 
The career glyph uses a sunburst structure with two levels of hierarchy to show the career sequence event distribution. 
The inner ring shows three title ranks (\ie, junior, intermediate, and senior) with purple in three saturation categories (Fig.~\ref{fig:system-ui}-C). 
The outer ring encodes five citation ranks in black at five saturation categories. 
The domain glyph is a doughnut chart that records the distribution of individuals' most frequently published paper venue types in the cluster. 
We highlight the visualization venues as pink and others as grey to show visualization researchers' domain diversity (Fig.~\ref{fig:system-ui}-C). 
The sector glyph is also a doughnut chart showing the social sector (\ie, academia, industry, and government) distribution: green for the industry, blue for the academia, and yellow for the government agency (Fig.~\ref{fig:system-ui}-C). 
The histogram on the right shows the citation distribution of the following year (\ie, DVs) of the sequence window. 
Users can compare the distribution of different clusters using the sequence list and choose a cluster of interest for further analysis. 

\textit{Justification}: 
Before designing three glyphs, we tried the proportional stacked area graph at first. 
For example, in Fig.~\ref{fig:visual-design-2-glyph}-D, we used the same colors to represent three title ranks. 
We used a navigator on the left to show the citation ranks within each title rank. 
However, it was space-wasting and hard to be integrated with other career information. 
Thus, we used glyphs to summarize multiple information of a group which could be reused in \textit{\comparisonviewname} and \textit{\personviewname}. 

\subsection{\personviewname} 
\wyf{After choosing a cluster from the \textit{\comparisonviewname}}, we use a \wyf{\textit{\careerlinename} to visualize an individual's careers and the effects of factors with} folded and unfolded modes in the \textit{\personviewname} (Fig.~\ref{fig:system-ui}-C, \textbf{T5}, \textbf{T6}). 

\begin{figure} [!htb]
 \centering 
 \vspace{-0.2cm}
 \includegraphics[width=\linewidth]{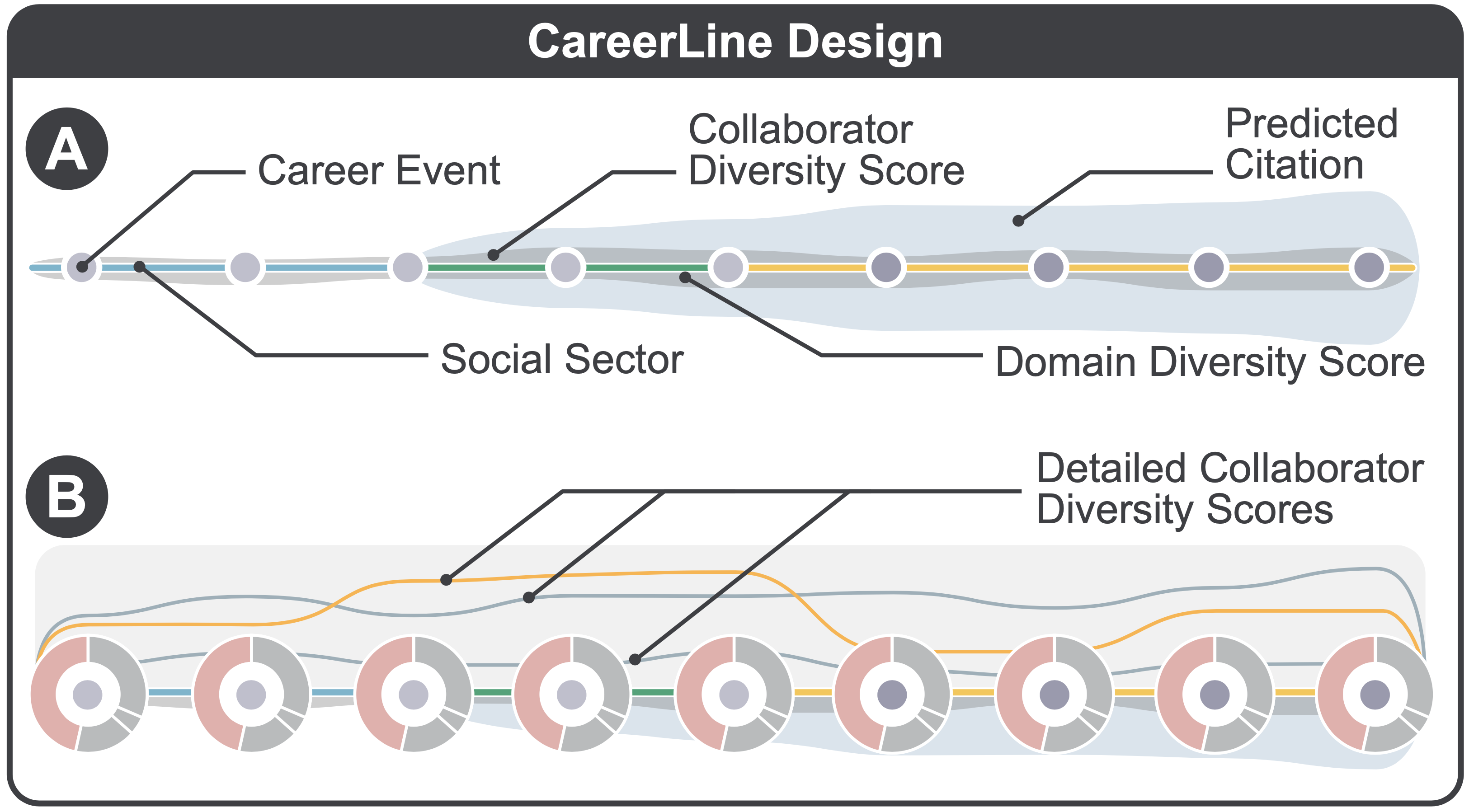}
 \vspace{-0.5cm}
 \caption{The visual design of \textit{\careerlinename} with folded and unfolded modes. 
 In the unfolded mode, users can have a detailed inspection of three collaborator diversity scores (i.e., career, domain, and sector diversities).}  
 \vspace{-0.4cm}
 \label{fig:visual-design-3-personline}
\end{figure}

\textit{Description}: 
In the folded mode (Fig.~\ref{fig:visual-design-3-personline}-A), the color of each circle shows the career title rank and the middle strip of the \textit{\careerlinename} represents the social sectors of the researcher, all with the same encoding in glyphs. 
Two flat gray areas distributed above and below the sector strip depict the collaborator and domain diversity scores of the researcher, respectively. 
The outer light blue area shows the predicted citations of the researcher based on the regression model. 
When hovering on a circle, a tooltip summarizing the researcher's domain diversity is shown (Fig.~\ref{fig:system-ui}-C4). 
The outer ring is a doughnut chart summarizing the paper distribution in different domains. 
\wyf{The inner word cloud is generated based on the paper venue names from the bibliographic data. 
We break the venue names into separate words and count the frequency statistics of each word to reflect the research topics. 
The word size encodes the number of papers.} 
%
Users can unfold the area of collaborator diversity score to study details about the researcher's collaborators (Fig.~\ref{fig:visual-design-3-personline}-B). 
A multi-line graph shows the collaborators' different diversity scores (\ie, career, domain, and sector). 
They can click one of the lines (\ie, diversities) to show the summary of the collaborators' population distribution via corresponding glyphs (Fig.~\ref{fig:visual-design-2-glyph}-A, B, C). 
We have provided sorting and filtering for users to choose researchers of interest. 
Users can rank them by average predicted citations or the collaborator diversity scores. 

\textit{Justification}: 
We tried to reveal all the collaborators' career paths as curves around the target researcher, where the distance of the two career circles is proportional to the number of papers they co-authored each year. 
However, it caused severe visual clutters when the number of coauthors increased. 
\wyf{In addition, according to our expert's feedback, showing the raw career paths of collaborators was useless for comparing individuals and finding drivers that may affect the target researcher's career success.} 
Thus, we computed different diversity scores to summarize the information of collaborators at each timestamp. 

\section{Evaluation}
\label{sec:06_Evaluation}
We evaluate the effectiveness and usability of the system using two case studies in Section~\ref{sec:06_CaseStudy} and interviews in Section~\ref{sec:06_ExpertInterview}. 

\subsection{Case Study}
\label{sec:06_CaseStudy} 
We invited our expert in Section~\ref{sec:03_Background_Concepts} to freely explore \textit{\systemname}. 
We then summarize the observations and comments and form them into two cases to fully demonstrate the system. 

\begin{figure*} [!htb]
 \centering 
 \vspace{-0.5cm}
 \includegraphics[width=\linewidth]{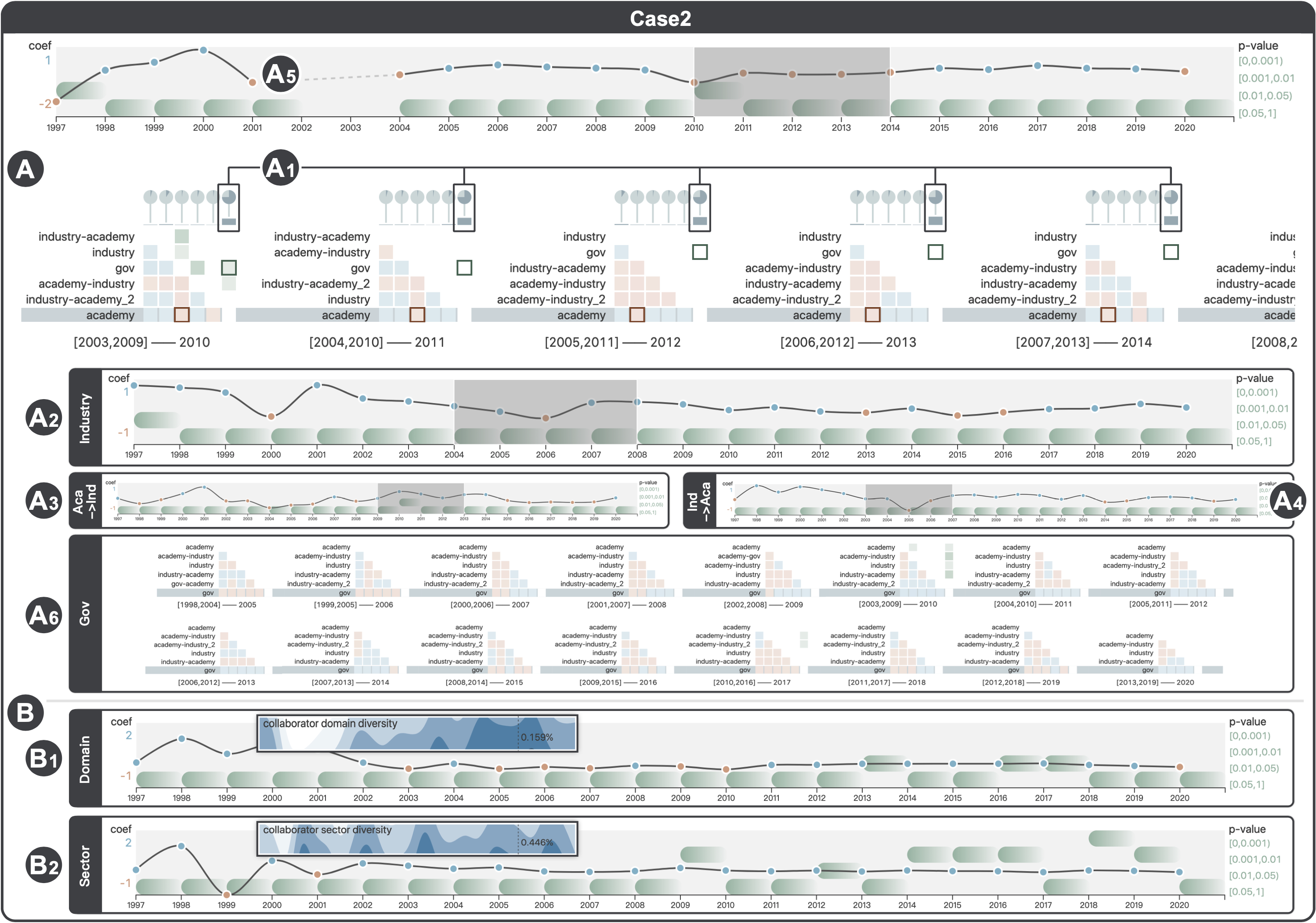} 
 \vspace{-0.6cm}
 \caption{The comparative study of (A) individual social sector (e.g., academia, industry, and government agency) and (B) collaborator diversity effects.} 
 \vspace{-0.6cm}
 \label{fig:case2}
\end{figure*} 

\subsubsection{Case1: Self-effort or Leverage?} 
\label{sec:06_Case1} 
Experts in \wyf{social science} have been engaged in studying multi-factor effects on academic career success for a long time. 

\textbf{Inter-factor-level Inspection.} 
{\ea} began with the \textit{\horizonchartname} (Fig.~\ref{fig:system-ui}-A) to obtain an overview of the impacts of all factors (\textbf{T1}, \textbf{T2}). 
\wyf{Hovering on the horizon charts, he found that in an overall sense individual factors outperformed social factors (\textbf{T2}). 
Nevertheless,} the individual factors (Fig.~\ref{fig:system-ui}-A1) had a great impact in the early era and gradually declined. 
On the contrary, social factors (Fig.~\ref{fig:system-ui}-A2), though did not contribute much at first, had become increasingly important (\textbf{T1}). 
\wyf{
\textit{``It has been a long-standing concern in social science whether one's career success should be attributed to human capital (\ie, individual factors) or social capital (\ie, social factors). 
With the booming of interdisciplinary collaborations, the role of social factors will become increasingly important. 
Nevertheless, the human capital still plays a pivotal role in their acquisition and accumulation of social capital, which validates that individual factors are still dominant factors.''}} 

\textbf{Intra-factor-level analysis.} 
{\ea} noticed that the first individual factor (i.e., career) outperformed the other two individual factors (Fig.~\ref{fig:system-ui}-A1). 
Wondering how different types of historical careers affected their upcoming career success (\textbf{T3}, \textbf{T4}), he chose this factor and turned to the \textit{\matrixchartname} (Fig.~\ref{fig:system-ui}-A3, A4). 
From the upper triangles, the pairwise comparisons among all categories were mostly statistically significant (green). 
\wyf{\textit{``...the differences among career sequence clusters are important in both statistical and substantial sense.''}} 
Viewing the \textit{\matrixtimelinename}, he found that senior researchers with a high citation rank (\ie, S/3 in Fig.~\ref{fig:system-ui}-A5) existed across most of the periods. 
\wyf{He clicked to align S/3 across time. 
From the \textit{\matrixtimelinename}, other clusters all performed worse than it (red cells) to affect career performance. 
It indicated that being a senior with high historical citations had obvious advantages on their upcoming career success. 
\textit{``This is related to the accumulated citation impacts of their previous works.''} 
Then, he set S/3 as the \textit{reference category} and chose other clusters (e.g., J-I/2 (Fig.~\ref{fig:system-ui}-A3), J-I/0-1 (Fig.~\ref{fig:system-ui}-A6), and S/1-2 (Fig.~\ref{fig:system-ui}-A7)) as targets to learn the impact evolvement.} 
From the \textit{\navigationtimelinename}, the temporal trend of each cluster's impact had not changed substantially. 
\textit{``Researchers' historical careers always have a strong and stable effect on their upcoming career performances. 
It has supported the view that one has to fight for themselves.''} 

{\ea} also wondered how the same sequence type (\ie, careers) in the social factor would affect researchers' careers. 
He thus set the \textit{collaborators' career types} as the target factor. 
In the \textit{\matrixtimelinename}, the S/3 group had a strong positive effect on researchers' career performance in the second window (Fig.~\ref{fig:system-ui}-A9). 
Thus, he aligned this cluster and went to the \textit{\navigationtimelinename} (Fig.~\ref{fig:system-ui}-A8). 
This group of collaborators negatively affected career performance at first (\ie, before 2010) and began to positively influence the career after around 2011. 
In addition, three windows (\ie, (1991, 1997), (1997, 2003), and (1998, 2004)) held a rare positive effect before 2010. 
{\ea} added these three clusters into the \textit{\comparisonviewname} for further analysis. 
\textit{``Interesting...Working with well-established researchers may not influence researchers' career performance immediately. 
They are usually working on innovative ideas, and the research outputs may take time to be seen by the world, which is always regarded as the `sleeping beauties in science'~\cite{ke2015defining}.''} 

\textbf{Individual-level Analysis.} 
After having an overview of the whole population, {\ea} wondered if individuals from a similar starting point would differ in their career success at the later stages (\textbf{T5}, \textbf{T6}). 
Turning to the \textit{\comparisonviewname} (Fig.~\ref{fig:system-ui}-B), he found that the third cluster (\ie, (1998, 2004)) was distinct (Fig.~\ref{fig:system-ui}-B1) due to the high homogeneity of career distribution and the high citations in the upcoming year (\ie, 2005). 
Thus, he chose it and went to the \textit{\personviewname}. 
\wyf{He ranked researchers by average predicted citations. 
The predicted citations of most researchers in the cluster displayed a similar growth trend during (1998, 2004) with a slight rise later. 
However, the first researcher ($R_{1}$) acted differently with a citation surge in the last several years (Fig.~\ref{fig:system-ui}-C1). 
$R_{1}$ also had a higher collaboration diversity score compared with others. 
Thus, {\ea} unfolded his \textit{\careerlinename}. 
$R_{1}$ had been collaborating with scholars from different sectors (Fig.~\ref{fig:system-ui}-C1). 
Hovering on career dots (Fig.~\ref{fig:system-ui}-C4), {\ea} found that $R_{1}$ changed research topics from robotics to vision. 
{\ea} also checked the collaboration and domain diversities of the first other ten researchers as representatives. 
Most of them had low diversities in several aspects (\eg, mostly published papers within VIS or working with those in academia such as Fig.~\ref{fig:system-ui}-C2, C3).} 
\wyf{\textit{``Individuals in the same cluster can differ from one another later, which is known as `within-group variance' in social science. 
$R_{1}$ is a case of the Matthew effect~\cite{merton1988matthew, feichtinger2021matthew}. 
Different strategies may cause diverse career performances even with the same starting point. 
For $R_{1}$, pursuing trendy topics and seeking diverse collaborations were potentially essential to improve his career.''}} 

The case demonstrated that \textit{\systemname} could help experts explore the dynamic multi-factor effects on academic career success at different levels of detail effectively. 

\subsubsection{Case2: Comparative Study} 
\label{sec:06_Case2} 
Comparison studies are commonly applied in \wyf{social science}. 
After exploring multiple factors, {\ea} found several stories summarized below. 

\textbf{Social sectors diversity effects.} 
{\ea} wondered whether historical sectors would influence career success (\textbf{T3, T4}). 
He thus chose the \textit{sector} factor and observed the \textit{\matrixtimelinename} (Fig.~\ref{fig:case2}-A). 
The sector clusters were stable across different windows and the academia cluster had been in the largest proportion (Fig.~\ref{fig:case2}-A1). 
He thus specified this cluster as the \textit{reference category} and compared it with industry-related clusters (\ie, industry, academia-industry, and industry-academia, Fig.~\ref{fig:case2}-A2, A3, A4). 
Generally, the career performance of individuals with industry experience performed better than those in academia alone. 
{\ea} inferred that it might be due to the imminent analytical needs in the industry that facilitate career performances. 
\textit{``Such borderless career movements could be a decent option for researchers' development.''} 
Temporally, he also noticed that three coefficient curves in the \textit{\navigationtimelinename} all suffered a decline before 2005,  \wyf{then the values all started to increase}. 
After obtaining the history of the VIS field, he learned that these special turning points might be related to the start of the VAST conference, which required cross-sector collaborations. 

For comparison, he highlighted those who had been working in the government agencies as the target category. 
From the coefficient curve (Fig.~\ref{fig:case2}-A5), the career performance presented a rare periodic pattern (\ie, rotating positive and negative effects every four or five years) compared with those staying in academia. 
\wyf{{\ea} further aligned this cluster to have a panoramic view by comparing it with all other clusters. 
From the \textit{\matrixtimelinename} (Fig~\ref{fig:case2}-A6), this cluster also presented a periodic pattern. 
It outperformed most clusters to affect the career performance in certain periods (\eg, 2005 to 2006, and 2015 to 2019) and was less important in other periods (\eg, 2010 to 2013).} 
{\ea} checked different window lengths and was surprised at the similar results. 
\textit{``I had never expected this before. 
It is possible that political policies may also affect the career performance of those in government agencies.''} 

\textbf{Collaborator diversity effects.} 
{\ea} wanted to learn how the diversities of categories in each social factor could facilitate one's career success (\textbf{T1, T2}). 
Thus, he compared three social factors which describe the number of categories in each factor (Fig.~\ref{fig:case2}-B). 
Interestingly, the diversity of collaborators' historical domain sequences (Fig.~\ref{fig:case2}-B1) had a negative effect at first and turned to be positive after around 2010. 
\textit{``Interdisciplinary collaborations always take time,''} {\ea} explained, \textit{``the collaborations at the early stage are always exploratory. 
Only after the trials will substantive collaborations appear and have a positive effect.''} 
For the collaborators' historical sector diversity (Fig.~\ref{fig:case2}-B2), it appeared to be almost all positive results over time. 
\textit{``It has cross-validated our previous findings. 
Researchers with more cross-border collaborators could outperform those working with collaborators in a single sector.''} 

The case showed that \textit{\systemname} could support comparative studies of multiple factors on academic career performance. 

\subsection{Expert Interview}
\label{sec:06_ExpertInterview} 
We first interviewed our domain expert {\ea}. 
Although the initial purpose of the system is for \wyf{social scientists} to explore the dynamic multi-factor effect on academic career success, the dataset we use may also be of interest to visualization (\ie, VIS) researchers for their career developments. 
Thus, we also invited four VIS researchers ({\visa}-{\visd}) who had not used \textit{\systemname} before. 
{\visa} and {\visd} are third-year PhD students with three-year VIS experience. 
{\visc} is a researcher with six-year of expertise. 
{\visb} just started her research as a first-year PhD student. 

\textbf{Procedure.} 
Each interview for VIS researchers lasted about 60 minutes. 
We first introduced the background of the project (\eg, the research problem, the data, and the analytical tasks). 
Second, we used a comprehensive example to demonstrate visual encoding and interactions. 
Third, we introduced the cases in Section~\ref{sec:06_CaseStudy} to show the insights and the usefulness of the system. 
Fourth, they could freely explore the system in a think-aloud manner. 
Finally, we conducted a post-study interview to ask for suggestions on system workflow, analytical framework, and visualization and interactions. 
We recorded their comments and findings during the process. 
The feedback of the two groups of users is summarized into three categories. 

\wyf{\textbf{System Workflow.} 
Two groups of users all considered the system workflow clear. 
{\ea} commented that \textit{\systemname} followed and strengthened their traditional analytical workflow, 
\textit{``it provides a panoramic overview to compare the impacts of multiple factors, which makes the analytical findings more valid and straightforward.''} 
He would follow the system logic to get familiar with \textit{\systemname} from factor comparison to individual inspection. 
Then he would focus on analyzing the regression results (\eg, \textit{\groupviewname} and \textit{\comparisonviewname}). 
The individual level was for case illustration and verification. 
For VIS researchers, those with statistical background (\ie, {\visc}) reacted similarly to {\ea}. 
Most insights he found were from the \textit{\groupviewname}. 
However, others (\ie, {\visa}, {\visb}, and {\visd}) found it hard to understand the matrix encoding and mostly used the \textit{\matrixtimelinename} to choose clusters and explored from the individual level. 
They would conclude with the evolving patterns of careers of specific researchers. 
VIS researchers also had different analytical foci compared with {\ea}. 
They preferred to filter based on attributes (\eg, choosing collaborators from a specific domain) instead of clustering. 
They also wanted more information such as the real names of researchers and institutions to find role models. 
In general, there was a barrier for non-experts (\ie, VIS researchers) to explore \textit{\systemname} due to diverse analytical foci and the lack of statistical background. 
It reflected that the target users of \textit{\systemname} were still social scientists. 
} 

\textbf{Dynamic Multi-factor Impact Analysis.} 
{\ea} appreciated the enhancement we added to the traditional SHA model, which made the whole framework more efficient. 
\textit{``The analytical approaches in social studies always follow a multi-step recipe. 
However, each step is troublesome that requires a lot of detailed processing semi-automatically. 
This automated framework indeed eases our burden.''} 
He was particularly impressed by the sequence slicing and cluster alignment. 
It helped him analyze the impact from a dynamic perspective, which he had never tried before. 
He also suggested making the lengths of sliding windows more flexible to capture both long-term and short-term impacts. 

\textbf{Visualization and Interactions.} 
All the users regarded the system as comprehensive to study the academic careers, which fulfilled all the analytical tasks. 
Most users liked the horizon charts, which gave a compact summary of multi-factors' dynamic impacts. 
{\ea} particularly liked the \textit{\matrixchartname}, \textit{``each matrix is with the same form as the table of coefficients and p-values we obtained from the R program, but in a more intuitive visual representation. 
The \textit{\matrixtimelinename} and the animation to align clusters across time are creative.''} 
For the \textit{\careerlinename} design, they could not remember all the design components at once. 
For example, at first, {\ea} mistook the two gray areas for the same diversity score in a symmetrical manner, which actually represented two types of diversities. 
{\visa} and {\visd} could not fully understand the matrix encoding since they were unfamiliar with the regression model. 
Nevertheless, after our explanation and exploring for a while, they finally got the point. 
All the users commented that most of the interactions were intuitive and straightforward. 
\wyf{{\ea} appreciated the well-coordinated system with multiple views displayed, 
\textit{``I can brush and study the data across views conveniently.''}} 
{\ea} and {\visc} suggested improving the system readability, such as adding annotations for visual designs and abbreviations. 
{\visb} wanted more interpretations related to the regression model. 


\section{Discussion}
\label{sec:07_Discussion} 
This section summarizes the significance of our work, the lessons we learned during our collaboration with social scientists, and the limitations and generalizations of \textit{\systemname}. 

\textbf{Significance.} 
In today's boundaryless career world, researchers' career choices have been increasingly flexible with more potential factors that may affect career success. 
We believe that the multi-factor impacts on academic career success will benefit both individuals' careers and the emerging literature in Science of Science. 
Traditional sociological methods are limited to analyzing from a static perspective. 
Therefore, we work closely with a social scientist to formulate the analytical tasks and perform the analysis longitudinally. 
\wyf{Our sequential explication of trajectories, integration of different factors, MIA model, and the visual analytical framework can also provide a new attempt for social scientists to study multi-factor impacts from a dynamic perspective beyond career studies (\eg, mobile news consumption~\cite{zhang2017structurally}).} 

\textbf{Lessons Learned.} 
The most important lesson is to take strengths from both domain and visualization fields to solve the problem more efficiently. 
\wyf{For the analytical model, we contributed to the sociological framework by adding the sliding window and cluster alignment to transform the traditional static analysis into a dynamic one.} 
For the visual design, \textit{\matrixtimelinename} is also derived from the table-based output from the R program and strengthened with new visualization and interactions (\eg, alignment). 
\wyf{
\wyfroundtwo{In summary, we obtained initial forms of the algorithm and visual design (\eg, SHA and matrix) from classical sociological studies.} 
One crucial strategy to fill the analytical gap that previous methods cannot solve is to work with experts and decompose the problem into a list of limitations step by step. 
Then we can apply approaches from different subject areas (\eg, computer science or other social science disciplines) to solve each limitation.} 
\wyf{Moreover, from the interview, although our expert and general researchers all expressed strong interest in the same dataset, they presented different analytical foci. 
Social scientists aim to find aggregated patterns underlying human behaviors while general researchers prefer detailed `real' information. 
In addition to diverse analytical foci, background knowledge (e.g., regression) should also be considered when designing the system.} 

\textbf{Limitations.} 
Although our evaluation has demonstrated the usefulness of \textit{\systemname}, it still has limitations. 
The first one is the lack of large-scale researcher data. 
Currently, we have collected around 1100 VIS researchers to demonstrate the system \wyf{due to the time-consuming data preparation process}. 
However, to support a comprehensive understanding of dynamic multi-factor effects on academic career success, larger datasets with a broader range are necessary. 
We plan to collect data from computer science or the whole science field to find more stories. 
Second, the fixed size of sliding windows smooths the variations in different timespans and may ignore short-term influences, which is also of interest to experts. 
We plan to dynamically set the window size based on the sequential trend to address this limitation. 

\textbf{Generalizability.} 
\wyf{First, \wyfroundtwo{the method for operationalizing factors can be generalized, especially for the impact analysis of factor histories. 
Users can apply MIA to distill typical factor history patterns and analyze the impacts in other domains such as finance and meteorology.} 
Particularly, to deal with the network nature, we have provided initial attempts to measure the social factors in multiple ways (Section~\ref{sec:04_Algorithm_MIA}). 
It can be adapted in other \wyfroundtwo{network impact} analysis such as the social relation impact on idea innovation~\cite{kijkuit2010little}. 
Second, visualization and interactions of showing the post-hoc tests temporally could be used in other pairwise comparisons of impact evolvement~\cite{peng2020filled}. 
The \textit{\careerlinename} (glyph + line chart) also provides new visual representations to show multivariate sequences with state transitions.} 
Third, \textit{\systemname} can be applied to other historical trajectory impact analysis using sequential data. 
The most similar one is to analyze careers beyond academia, such as human resources (HR) management in companies. 
HR professionals can employ it to provide individualized trainings for employees to enhance their performance. 
A border usage is to analyze other life-course data (\eg, health and migration~\cite{christakis2007spread}) to understand human behaviors. 
Before applying the system, we suggest practitioners first organize different factors into multiple sequences. 
Then they can specify a target sequence type (\eg, career success in our study) to analyze. 

\section{Conclusion}
\label{sec:08_Conclusion} 
This paper has presented \textit{\systemname}, an interactive visualization system that enables \wyf{social scientists} to explore the multi-factor impacts on academic career success from a dynamic perspective. 
It supports the analysis from three levels of detail, including inter-factor level comparison, intra-factor level exploration, and individual level inspection. 
We have proposed two novel visual designs (i.e., \textit{\matrixchartname} and \textit{\careerlinename}) to show the factor impacts. 
Two case studies and the interviews have shown the usefulness of the system. 

In the future, we plan to first enhance the current MIA framework by setting dynamic lengths of sliding windows based on sequential trends. 
Second, we aim to collect data from a border range (\eg, computer science field) to find more stories. 
\wyf{Third, we plan to gather requirements from general researchers to adapt \textit{\systemname} for wider use.} 

\balance

\end{spacing}



\maketitle

\vspace{+0.1cm}
\acknowledgments{
This research was supported in part by Hong Kong Theme-based Research Scheme grant T44-707/16-N, GRF 11505119 from HKSAR Research Grants Council, NSFC (62072400), Zhejiang Provincial Natural Science Foundation (LR18F020001), and the Collaborative Innovation Center of Artificial Intelligence by MOE and Zhejiang Provincial Government (ZJU).  
}

\bibliographystyle{abbrv-doi}

\newpage
\bibliography{main}
\end{document}